\documentclass[a4paper, 11pt]{article}

\usepackage{graphicx}
\usepackage[all]{xy}
\usepackage{amsfonts}
\usepackage{dsfont}
\usepackage{mathrsfs}
\usepackage{amssymb}
\usepackage{bbm}
\usepackage{epsfig}
\usepackage{amsmath}
\usepackage{appendix}
\usepackage{color}
\usepackage[english]{babel}

\def\Reals{\mathop{\hbox{\mit I\kern-.2em R}}\nolimits}

\def\Complexes{{\hbox{\mit C\kern-.46em
            \vrule depth 0ex height 1.4ex width .05em\kern.41em}}}

\usepackage[english]{babel}
\usepackage{epsfig}
\usepackage{dsfont}
\usepackage{mathrsfs}
\usepackage[mathscr]{eucal}
\usepackage{graphicx}
\usepackage{amsmath,amsfonts,amsthm,amssymb,amsbsy,amsopn,amstext}
\usepackage{latexsym}
\usepackage{color,multicol,makeidx}
\usepackage{CJK}
\usepackage{indentfirst}
\usepackage{bm}
\usepackage{mathrsfs}
\usepackage{float}
\usepackage[caption=false,font=normalsize,labelfont=sf,textfont=sf]{subfig}
\usepackage{multirow}

\newtheorem{thm}{Theorem}
\newtheorem{defn}{Definition}

\newtheorem{lem}{Lemma}
\newtheorem{remark}{Remark}
\newtheorem{prop}{Proposition}

\title{\bf Emergent Behaviors  over Signed Random Networks \\ in Dynamical Environments}
\date{}
\author{Guodong Shi, Alexandre Proutiere, Mikael  Johansson, John. S. Baras, and Karl H. Johansson\thanks{G. Shi, A. Proutiere, M.  Johansson, and K. H. Johansson are with ACCESS Linnaeus Centre, Royal Institute of Technology, Stockholm 10044, Sweden. J. S. Baras is with Electrical and Computer Engineering, University of Maryland, College Park, MD 20742, USA. e-mail: guodongs@kth.se, alepro@kth.se, mikaelj@kth.se, baras@umd.edu, kallej@kth.se. This work has been supported in part by the Knut   and Alice Wallenberg Foundation, the Swedish Research Council,   KTH SRA TNG, and  by AFOSR MURI grant FA9550-10-1-0573.}}
\begin{document}
\maketitle

%%%%%%%%%%%%%%%%%%%%%%%%%%%%%%%%%%%%%%%%%%%%%%%%%%%%%%%%%%%%%%%%%%%%%%%%%%%%%%%%%%%%
\begin{abstract}
  We study asymptotic dynamical patterns that emerge among a set of nodes that interact in a
dynamically evolving signed random network.  Node  interactions take place at random on a sequence of deterministic signed graphs.  Each node receives positive or negative  recommendations from its neighbors depending on the sign of the interaction arcs, and updates its state accordingly. Positive recommendations follow the standard consensus update while two types of negative recommendations, each modeling a different type of antagonistic or malicious interaction, are considered. Nodes may weigh positive and negative recommendations differently, and random processes are introduced to model the time-varying attention that nodes pay to the positive and negative recommendations.  Various conditions for almost sure convergence, divergence, and clustering of the node states are established. Some fundamental similarities and differences  are established  for the two notions of negative recommendations.
\end{abstract}

{\bf Keywords.} Random graphs, Signed networks, Consensus dynamics, Belief clustering

%{\bf Keywords:} Consensus algorithms, , State agreement, State polarization
\section{Introduction}
\subsection{Motivation}
The need to model, analyze and engineer large complex networks appears  in a wide range of scientific disciplines, ranging from social sciences and biology to physics and engineering \cite{degroot,vic95,jad03}. In many cases, these networks are composed of relatively simple agents that interact locally with their neighbors based on a very limited knowledge about the system state.
Despite the simple local interactions, the resulting networks can display a rich set of emergent behaviors, including certain forms of intelligence and learning~\cite{swarm,naive}.

%Despite the simple local interactions, the resulting networks can display very complex dynamics, and the global behavior sometimes be very sensitive to the parameters that define the interactions.

Consensus problems, in which the aim is to compute a weighted average of the initial values held by a collection of nodes, play a fundamental role in the study of node dynamics over complex networks.   Early  work \cite{degroot} focused on understanding how opinions evolve in a network of agents, and showed that a simple deterministic opinion update based on the mutual trust and the differences in belief between interacting agents could lead to global convergence of the beliefs. Consensus dynamics has since then been widely adopted for describing opinion dynamics in social networks, e.g., \cite{naive,social2,misinformation}. In engineering sciences, a huge amount of literature has studied these algorithms for distributed averaging, formation  forming and load balancing between collaborative agents under fixed or time-varying interaction networks \cite{tsi,xiao,juliencdc,mor,ren,saber,caoming,julien13}. Randomized  consensus seeking has also been widely studied, motivated by  the random nature of interactions and updates in real complex networks~\cite{hatano,boyd,fagnani,kar2,touri,jad08,it10,TON13}.

Interactions in large-scale networks are not always collaborative, as nodes often take on different, or even opposing, roles. A convenient framework for modeling different roles and relationships between agents is to use signed graphs. Signed graphs were introduced in the classical work by Heider in 1940s~\cite{heider46} to model the structure of social networks, where a positive link represents a friendly relation between two agents, and a negative link an unfriendly one.  In \cite{galam96}, a dynamic model based on a signed graph with positive links between nodes (representing nations) belonging to the same coalition and negative otherwise,  was introduced to study  the stability of  world politics. In biology, sign patterns have been used to describe activator--inhibitor interactions between pairs of chemicals \cite{biologymath},  neural networks for vision and learning \cite{adptive}, and gene regulatory networks~\cite{nature13}. In all these examples, the state updates that happen when two nodes interact depend on the sign of the arc between the nodes in the underlying graph. The understanding of the emergent dynamical behaviors of networks with agents having different roles is much more limited than for instance collaborative agents performing consensus algorithms.

%many networks include hostile or antagonistic nodes. These can represent enemies in a social network, or a malicious or compromised device in a computer network.
% The signed graph has a vertex for every agent and a link for each pair of agents that can interact. Each arc is associated with a sign which encodes whether the nodes are friendly or hostile, and the behavior of interacting nodes then depend on the sign of the arc. Multi-agent interactions on such signed networks can generate a large number of phenomena observed in real-world social networks, such as agreement, fragmentation, polarization, and extremism [reference] In addition, many real networks display a large heterogeneity in the frequency and intensity of interactions.[reference]

It is intriguing to investigate  what happens when two types of dynamics are coupled in a single network. Naturally we ask: how we should model the dynamics of positive and negative interactions, when do behaviors such as consensus, swarming and clustering emerge, and how does the structure of the sign patterns influence these behaviors? In this paper, we answer these questions for a general model of opinion formation in dynamic signed random networks.
% in a \textcolor{blue}{dynamic environment and answer some of}.
% We \textcolor{blue}{assume that positive interactions follow a standard consensus update and consider two types of negative interactions.  define the positive interactions as the standard %consensus dynamics, and  two types of negative interactions are discussed.}

\subsection{Contribution}

We consider  general  randomized node interactions.  A sequence of deterministic signed graphs defines a dynamical environment of the network, and then random node interactions take place under independent,  but not necessarily identically distributed,  random sampling of the environment.  Once interaction relations have been realized, each node receives a positive recommendation consistent with the standard consensus algorithm from its positive neighbors. Nodes receive negative recommendations from the negative neighbors. Following \cite{altafini13,shiJSAC} we consider two models of negative recommendations. In the state-reversion recommendation,  each node receives false values from its negative neighbors without necessarily  knowing which of its neighbors is positive or negative \cite{altafini13}. In the relative-state-reversion model, nodes receive a repulsive influence from their known negative neighbors. After receiving these recommendations, each node puts a (deterministic) weight to each recommendation, and then encodes these weighted recommendations in its state update through (stochastic) attentions defined by two Bernoulli random variables.

We establish conditions for the almost sure convergence, divergence, and clustering of the node states for the considered signed random networks. Fundamental similarities and differences  are established  for the two models of negative recommendations. We show that strong structural balance~\cite{harary56} is crucial for belief clustering in the state-reversion model (which is consistent with the work of Altafini \cite{altafini13}), while weak structural balance is enough in the relative-state-reversion model. We also show that the deterministic weight and the stochastic attention nodes put on recommendations play a drastically different role for the state convergence and divergence of the network in the two models. The models  share some consistent behaviors: if some arc independence is imposed on the random interactions, two similar no-survivor statements are established, which generalize the results  for the gossiping model in \cite{shiOR}.

% The interaction model accounts for both frequency and intensity of interactions, and includes and extends the classical DeGroot model for opinion dynamics, Altafini's model of networks for consensus in networks with antagonistic links, as well as our earlier work on opinion formation in signed social networks

\subsection{Paper Organization}
In Section 2 we propose the network dynamics and the node update rules. State-reversion and relative-state-reversion models are proposed, respectively, for the negative recommendations of the node updates. Section 3 presents our main results on the state-reversion model as well as the proofs. Then Section 4 moves to the relative-state-reversion model and finally some concluding remarks are drawn in Section 5.
\vspace{3mm}
\subsection*{Graph Theory, Notations and Terminologies}

A simple directed graph (digraph) $\mathcal
{G}=(\mathcal {V}, \mathcal {E})$ consists of a finite set
$\mathcal{V}$ of nodes and an arc set
$\mathcal {E}\subseteq \mathcal{V}\times\mathcal{V}$, where  $e=(i,j)\in\mathcal {E}$ denotes   an
{\it arc}  from node $i\in \mathcal{V}$  to $j\in\mathcal{V}$ with $(i,i)\notin \mathcal{E}$ for all $i\in\mathcal{V}$.  We call node $j$  {\it reachable} from node $i$ if there is a directed path from $i$ to $j$. In particular every node is supposed to be reachable from itself. A node $v$ from which every  node in $\mathcal{V}$ is reachable is called a {\it center node} (root). A digraph $\mathcal{G}$ is {\it strongly connected} if every two  nodes are mutually reachable;  $\mathcal{G}$ has a spanning tree if it has a center node; $\mathcal{G}$ is {\it weakly connected} if  a connected undirected graph can be obtained by removing all the directions of the arcs in $\mathcal{E}$. A subgraph  $\mathcal
{G}$ of $\mathcal
{G}=(\mathcal {V}, \mathcal {E})$, is a graph on the same node set $\mathcal {V}$ whose arc set is a subset of $\mathcal {E}$. The induced graph of $\mathcal{V}_i \subseteq \mathcal{V}$ on $\mathcal{G}$, denoted $\mathcal{G}|\mathcal{V}_i$, is the graph $(\mathcal{V}_i, \mathcal{E}_i)$ with $\mathcal{E}_i=(\mathcal{V}_i\times \mathcal{V}_i)\cap \mathcal{E}$. A weakly connected component of $\mathcal
{G}$ is a maximal weakly connected induced graph of $\mathcal
{G}$. If each arc $(i,j)\in\mathcal{E}$ is associated uniquely with a sign, either '$+$' or '$-$', $\mathcal {G}$ is called a signed graph and the sign of $(i,j)\in\mathcal {E}$ is denoted  as $\sigma_{ij}$.  The positive and negative subgraphs containing the positive and negative arcs of $\mathcal{G}$, are denoted as $\mathcal {G}^{+}=(\mathcal{V}, \mathcal{E}^+)$ and $\mathcal {G}^-=(\mathcal{V}, \mathcal{E}^-)$, respectively.

 Depending on the argument, $|\cdot|$ stands for the absolute value of a real number, the Euclidean norm of a vector or the cardinality of a set. The $\sigma$-algebra of a random variable is denoted as $\sigma(\cdot)$.   We use $\mathds{P}(\cdot)$ to denote the probability and $\mathds{E}\{\cdot\}$ the expectation
 %and  $\mathds{V}\{\cdot\}$ the variance
 of their arguments, respectively.

%A finite square matrix $M=[m_{ij}]\in\mathbb{R}^{n\times n}$ is called {\em stochastic} if $m_{ij}\geq 0$ for all $i,j$ and $\sum_j m_{ij}=1$ for all $i$.  A stochastic matrix $M$ is  %{\em doubly stochastic} if also $M^T$ is  stochastic. Let $P=[p_{ij}]\in\mathbb{R}^{n\times n}$ be a matrix with nonnegative entries. We can associate a unique digraph  $\mathcal {G}_P=(\mathcal{V},\mathcal{E}_P)$ with $P$ on node set $\mathcal{V}=\{1,\dots,n\}$ such that $(j,i)\in\mathcal{E}_P$ if and only if $p_{ij}>0$. We call $\mathcal {G}_P$ the {\em induced graph} of $P$.

\section{Networks Dynamics and Node Updates}

We consider a dynamic  network where each user holds and updates her  belief or  state when interacting with other users. In this section, we present a general model specifying the network dynamics and the way users interact.

%%%%%%%%%%%%%%%%%%%%%%%%%%%%%%%%%%%%%%%%%%%%%%%%%%%%%%%%%%%%%%%%%%%%%%%%%%%%%%%%%%%%%%%%%%%%%%%%%%%%%
\subsection{Dynamic Signed Graphs}

We consider a network  with a set $\mathcal{V}=\{1,\dots,n\}$ of $n$ users or nodes, with $n\geq3$. Time is slotted, and at each slot $t=0,1,\ldots$, each user can interact with her neighbors in a simple  directed graph $\mathcal {G}_t=(\mathcal{V},\mathcal{E}_t)$. The graph evolves over time in an arbitrary and deterministic manner. We assume ${\cal G}_t$ is a signed graph, and we denote by $\sigma_{ij}(t)$ the sign of arc $(i,j)\in \mathcal{E}_t$. The sign of arc $(i,j)$ indicates whether $i$ is a friend ($\sigma_{ij}(t)=+$), or an enemy ($\sigma_{ij}(t)=-$) of node $j$. The positive and negative subgraphs containing the positive and negative arcs of $\mathcal{G}_t$, are denoted by $\mathcal {G}^{+}_t=(\mathcal{V}, \mathcal{E}^+_t)$ and $\mathcal {G}^-_t=(\mathcal{V}, \mathcal{E}^-_t)$, respectively. We say that the sequence of graphs $\{\mathcal {G}_t\}_{t\ge 0}$ is {\it sign consistent} if the sign of any arc $(i,j)$ does not evolve over time, i.e., if for any $s,t \ge 0$,
$$
(i,j)\in \mathcal {E}_s\ {\rm and}\ (i,j)\in \mathcal {E}_t\  \Longrightarrow \sigma_{ij}(s)=\sigma_{ij}(t).
$$
We also define $\mathcal {G}_\ast=(\mathcal{V},\mathcal{E}_\ast)$ with $\mathcal{E}_\ast=\bigcup_{t=0}^\infty\mathcal{E}_t$ as the total graph of the network. If  $\{\mathcal {G}_t\}_{t\ge 0}$ is sign consistent, then the sign of  each arc $\mathcal{E}_\ast$ never changes and in that case, $\mathcal {G}_\ast=(\mathcal{V},\mathcal{E}_\ast)$ is a well-defined signed graph.

Next we introduce the notion of positive cluster in a signed directed graph (digraph), which will play an important role in the analysis of the belief dynamics.

\begin{defn} Let $\mathcal {G}$ be a signed digraph with positive subgraph $\mathcal {G}^+$.
 A subset $\mathcal {V}_\ast$ of the set of nodes $\mathcal {V}$ is a  positive cluster  if $\mathcal {V}_\ast$ constitutes a weakly connected component of $\mathcal {G}^{+}$. A positive cluster partition of $\mathcal {G}$ is a partition of $\mathcal{V}$ into $\mathcal{V}=\bigcup_{i=1}^{{\rm T}_{\rm p}} \mathcal{V}_i$ for some ${\rm T}_{\rm p}\geq 1$, where for all $i=1,\ldots, {\rm T}_{\rm p}$, $\mathcal{V}_i$ is a  positive cluster.
 \end{defn}
Note that negative arcs may exist between the nodes of a positive cluster. Therefore, $\mathcal {G}$ admitting a positive-cluster partition is a generalization of the classical definition of weakly structural balance for which negative links are strictly forbidden  \cite{davis63}.  From the above definition, it is clear that for any signed graph $\mathcal {G}$, there is a unique positive cluster  partition $\mathcal{V}=\bigcup_{i=1}^{{\rm T}_{\rm p}} \mathcal{V}_i$ of $\mathcal{G}$, where ${\rm T}_{\rm p}$ is the  number of positive clusters covering the entire set ${\cal V}$ of nodes.

\begin{figure}[t]
\begin{center}
\includegraphics[height=2.4in]{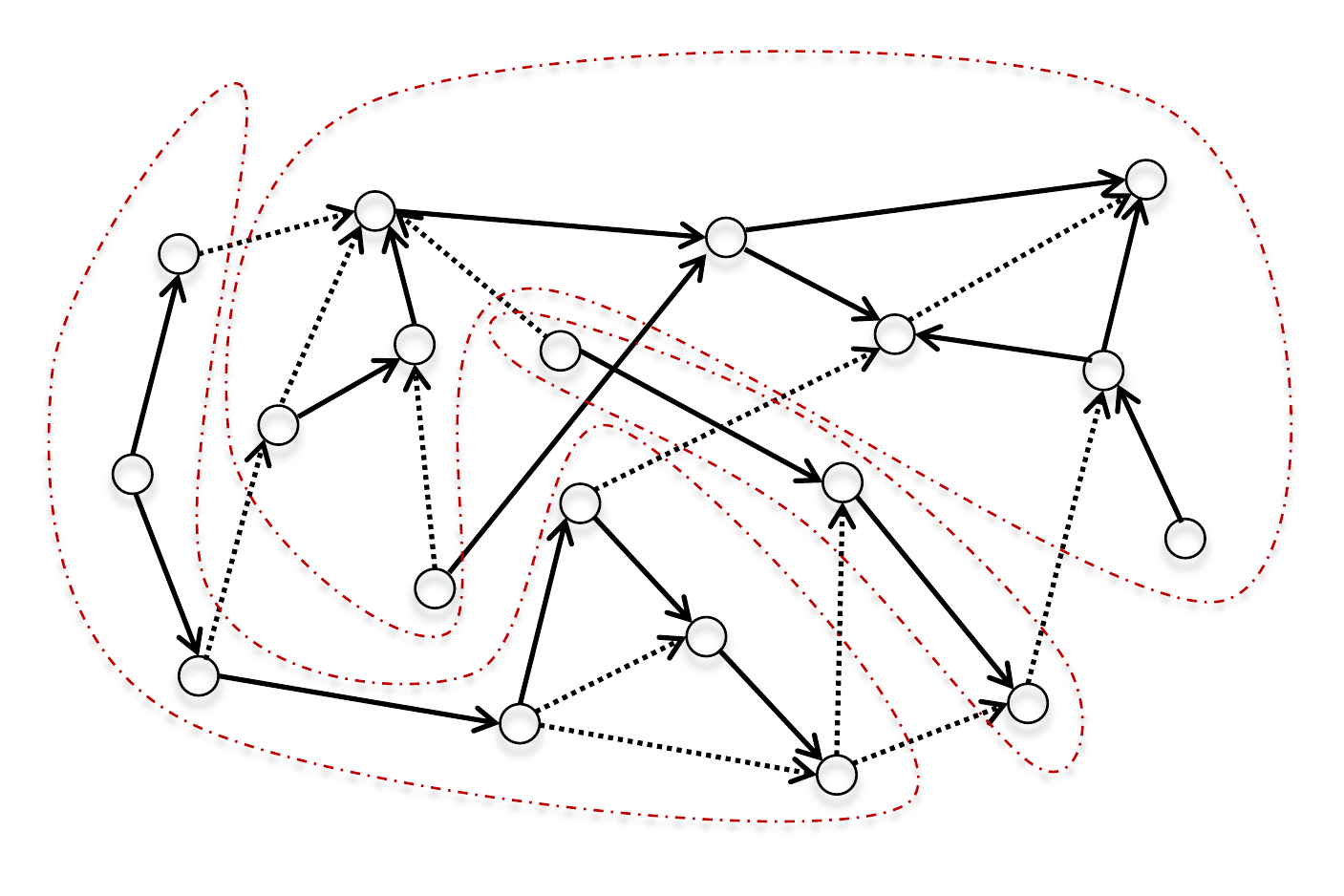}
\caption{A signed network and its three positive clusters. The  positive arcs are solid, and the negative arcs are dashed. Note that negative arcs are allowed within  positive clusters. } \label{frontier}
\end{center}
\end{figure}

%%%%%%%%%%%%%%%%%%%%%%%%%%%%%%%%%%%%%%%%%%%%%%%%%%%%%%%%%%%%%%%%%%%%%%%%%%%%%%%%%%%%%%%%%%%%%%%
\subsection{Random Interactions}

At time $t$, node $i$ may only interact with her neighboring nodes in ${\cal G}_t$. We present a general model on the random node interactions  at a given time $t$. This model includes the classical Erd\H{o}s-R\'{e}nyi random graph \cite{er}, gossiping models where a single pair of nodes is chosen at random for interaction \cite{boyd}, as well as where all nodes interact with their neighbors at a given time \cite{fagnani}. At time $t$, some pairs of nodes are randomly selected for interaction. We denote by $E_t\subset {\cal E}_t$ the random subset of arcs corresponding to interacting node pairs at time $t$. To be precise, $E_t$ is sampled from the distribution $\mu_t$ defined over the set $\Omega_t$ of all subsets of arcs in ${\cal E}_t$. We assume that $E_0,E_1,\ldots$ form a sequence of independent sets of arcs. Formally, we introduce the probability space $(\Theta,\mathcal{F},\mathrm{P})$ obtained by taking the product of the probability spaces $(\Omega_t,{\cal S}_t,\mu_t)$, where $\mathcal{S}_t$ is the discrete $\sigma$-algebra on $\Omega_t$: $\Theta=\prod_{t\ge 0}\Omega_t$, ${\cal F}$ is the product of $\sigma$-algebras ${\cal S}_t$, $t\ge 0$, and $\mathrm{P}$ is the product probability measure of $\mu_t, t\ge 0$. We  denote by $G_t=(\mathcal{V}, E_t)$ the random subgraph of ${\cal G}_t$ corresponding to the random set $E_t$ of arcs. The disjoint  sets $E^+_t$ and $E^-_t$ denote the positive and negative arc set of $E_t$, respectively. Finally, we split the random set of nodes interacting with node $i$ at time $t$ depending on the sign of the corresponding arc: for node $i$, the set of positive neighbors is defined as ${N}^+_i(t):= \big\{j: (j,i)\in E_t^+\big\}$, whereas similarly, the set of negative neighbors is ${N}^-_i(t):= \big\{j: (j,i)\in E_t^-\big\}$.

%%%%%%%%%%%%%%%%%%%%%%%%%%%%%%%%%%%%%%%%%%%%%%%%%%%%%%%%%%%%%%%%%%%%%%%%%%%%%%%%%%%%%%%%%%%%%%%
\subsection{Node updates}

Next we explain how nodes update their states. Each node $i$ holds a state $s_i(t)\in \mathds{R}$ at $t=0,1,\dots$. To update her state at time $t$, node $i$ considers recommendations received from her positive and negative neighbors:
\begin{itemize}
 \item[(i)] The positive  recommendation node $i$ receives  at time $t$ is
$h_i^+(t):=-\sum_{j\in{N}_i^+(t) } \big(s_i(t)-s_j(t)\big)$.
\item[(ii)] The negative recommendations node  $i$ receives at time $t$ are modeled by two different maps:
\begin{itemize}
\item The state-reversion   recommendation $h_i^-(t):=- \sum_{j\in{N}_i^-(t) } \big(s_i(t)+s_j(t)\big)$;
\item The relative-state-reversion  recommendation $h_i^-(t):= \sum_{j\in{N}_i^-(t) } \big(s_i(t)-s_j(t)\big)$.
\end{itemize}
\end{itemize}
In the above expressions, we use the convention  that summing over empty sets yields a recommendation equal to zero, e.g., when node $i$ has no positive neighbors, then $h_i^+(t)=0$.
\begin{remark}
The two definitions of negative recommendations have  different physical interpretations and make different assumptions on the knowledge that nodes possess  about their neighbor relationships.  The state-reversion model can be interpreted as a situation where negative nodes provide false values of their states by flipping the true sign~\cite{altafini13}. However, the receiving node does not necessarily know which of its neighbors are positive or negative. In the relative-state-reversion model, on the other hand, nodes must know if a specific neighbor is positive or negative to implement the state update that causes the  repulsive influence from its negative neighbors \cite{shiJSAC}.
\end{remark}

Now let $\{{B}_t\}_{t\ge 0}$ and $\{{D}_t\}_{t\ge 0}$ be two sequences of independent Bernoulli random variables. We assume that $\{{B}_t\}_{t\ge 0}$, $\{{D}_t\}_{t\ge 0}$, and $\{G_t\}_{t\ge 0}$ define independent processes. For any $t\ge 0$, define $b_t=\mathds{E}\{B_t\}$ and $d_t=\mathds{E}\{D_t\}$. The processes $\{{B}_t\}_{t\ge 0}$ and $\{{D}_t\}_{t\ge 0}$ represent how much  attention node $i$ pays to  the positive and negative recommendations, respectively.

Node $i$ updates her state as follows:
\begin{align}\label{9}
 s_i(t+1)=s_i(t)+  \alpha{B}_t h_i^+(t) +\beta{D}_t h_i^-(t),
\end{align}
where $\alpha,\beta>0$ are two positive constants marking the  weight each node put on the positive and negative recommendations, respectively. Depending on the definition of $h_i^-(t)$, we call the corresponding model the state-reversion model and the relative-state-reversion model, respectively.

Let  $s(t)=\big(s_1(t)\dots s_n(t)\big)^T$ be the random vector representing the network state  at time $t$. The main objective of this paper is to analyze the behavior of the stochastic process $\{s(t)\}_{t\ge 0}$. In the following, we denote by $\mathds{P}$ the probability measure capturing all random components driving the evolution of $s(t)$.

\section{The State-Reversion  Model}
In this section, we study the system dynamics under the state-reversion model. We provide conditions for convergence and divergence. The results are stated in the following subsection, and the remaining of the section is devoted to their proofs.

\subsection{Main Results}

We begin by stating two natural assumptions on the way nodes are selected for updates, and on the graph dynamics. In the first assumption, we impose that  at time $t$, any arc is selected in $E_t$ with positive probability. The second assumption states that the unions of the graphs ${\cal G}_t$ over time-windows of fixed duration are strongly connected.

\noindent {\bf A1.} There is a constant  $p_\ast \in (0,1)$ such that for all $t\geq0$ and  $i,j\in \mathcal{V}$, $\mathds{P}\big( (i,j)\in E_t \big)\geq p_\ast$ if $(i,j)\in \mathcal{E}_t$.

\noindent {\bf A2.} There is an integer $K \geq 1$ such that the union graph $\mathcal{G}\big({[t,t+K-1]}\big)=\big(\mathcal{V}, \bigcup_{\tau \in [t,t+K-1] } \mathcal{E}_\tau\big)$ is strongly connected for all $t\geq0$.

The following theorem provides conditions under which the system dynamics converges almost surely. Surprisingly, these conditions are mild: we just require that the sum of the updating parameters $\alpha$ and $\beta$ is small enough, and that node updates occur with constant probabilities, i.e., $\mathds{E}\{B_t\}$ and $\mathds{E}\{D_t\}$ do not evolve over time. In particular, the state of each node converges almost surely even if the signs of arcs change over time.

\begin{thm}\label{thmsr1}
Assume that A1 and A2 hold, and that $\alpha,\beta>0$ are such that $\alpha+\beta  \leq {1}/{(n-1)}$. Further assume that for any $t\ge 0$, $b_t\equiv b$ and $d_t\equiv d$ for some $b,d \in (0,1)$. Then under the state-reversion  model, we have, for all $i\in\mathcal{V}$ and all initial states $s(0)$,
$
\mathds{P} \big( \lim_{t\rightarrow \infty} s_i(t) \ {\rm exists}\big)=1.
$
\end{thm}

In the above theorem, we say that $\lim_{t\rightarrow \infty} s_i(t)$ exists if $s_i(t)$ converges to a finite limit as $t$ tends to infinity. Characterizing the limiting states  is in general challenging. There are however scenarios where this can be done, which require  the notion of structural balance~\cite{{harary56}}.

\begin{defn}
Let $\mathcal {G}=(\mathcal{V}, \mathcal{E})$ be a signed digraph. $\mathcal {G}$ is strongly balanced if  we can divide $\mathcal{V}$ into two disjoint  nonempty subsets $\mathcal{V}_1$ and $\mathcal{V}_2$ where negative arcs exist only between these two subsets.
\end{defn}

To predict the limiting system behavior, we make the following assumption.

\noindent
{\bf A3.}  $\{\mathcal{G}_t\}_{t\ge 0}$ is sign consistent admitting  a  total graph  $\mathcal{G}_\ast$.

\begin{thm}\label{thmsr2}
Assume that A1, A2 and A3 hold, and that $\alpha,\beta>0$ are such that $\alpha+\beta  \leq {1}/{(n-1)}$. We suppose $\mathcal{G}_\ast$ contains at least one negative arc and every negative arc in $\mathcal{G}_\ast$ appears infinitely often in $\{\mathcal{G}_t\}_{t\ge 0}$. Further assume that for any $t\ge 0$, $b_t\equiv b$ and $d_t\equiv d$ for some $b,d \in (0,1)$. Then under the state-reversion model, we have, for any initial state $s(0)$:

(i) If $\mathcal{G}_\ast$ is strongly balanced, then there is a random variable $y_\ast$, with $y_\ast \leq \|s(0)\|_1$ almost surely, such that
$
\mathds{P} \big( \lim_{t\rightarrow \infty} s_i(t) = y_\ast, \forall i\in \mathcal{V}_1; \  \lim_{t\rightarrow \infty} s_i(t) =- y_\ast, \forall i\in \mathcal{V}_2\big)=1;
$

(ii) If $\mathcal{G}_\ast$ is not strongly balanced, then
 $
\mathds{P} \big( \lim_{t\rightarrow \infty} s_i(t) = 0, \forall i\in \mathcal{V}\big)=1.
$
\end{thm}

Theorem \ref{thmsr2} states that strong structural balance is crucial to ensure convergence to nontrivial clustering states, which is consistent with the result of \cite{altafini13} derived for fixed graphs under continuous-time node updates. Instead of a spectral analysis as in \cite{altafini13}, we study the asymptotic behavior of each sample path. From the above theorem, we know that under the strong structural balance condition, the states of nodes in the same positive cluster converge to the same limit, and that the limits of two nodes in different positive clusters are exactly opposite. Using similar arguments as in \cite{altafini13}, the value of $y_\ast$ can be described as the limit of a random consensus process with the help of a gauge transformation.

Next we are interested in determining whether the states  could diverge depending on the values of the updating parameters $\alpha$ and $\beta$. We show that by increasing $\beta$, i.e., the strength of the negative recommendations,  one may observe such divergence. To this aim, we make the following assumptions.

\noindent {\bf A4.}  There is an integer $K\geq 1$ such that the union graph $\mathcal{G}^+\big({[t,t+K]}\big)=\big(\mathcal{V}, \bigcup_{\tau \in [t,t+K-1] } \mathcal{E}_\tau^+\big)$ is strongly connected for all $t\geq0$.

\noindent {\bf A5.}  There is an integer $K\geq 1$ such that the union graph $\mathcal{G}^-\big({[t,t+K]}\big)=\big(\mathcal{V}, \bigcup_{\tau \in [t,t+K-1] } \mathcal{E}_\tau^-\big)$ is strongly connected for all $t\geq0$.

\noindent {\bf A6.} The events  $\{ (i,j)\in G_t \}$, $i,j\in\mathcal{V}$, $t=0,1,\dots$ are independent and there is a constant  $p^\ast \in (0,1)$ such that for all $t\geq0$ and  $i,j\in \mathcal{V}$, $\mathds{P}\big( (i,j)\in G_t \big)\leq  p^\ast$ if $(i,j)\in \mathcal{E}_t$.

\begin{thm}\label{thmsr3}
Assume that A1, A4, A5 and A6 hold, and that for any $t\ge 0$, $b_t\equiv b$ and $d_t\equiv d$ for some $b,d \in (0,1)$. Fix $\alpha \in[0,(2n)^{-1}]$. Then under the state-reversion  model, there is $\beta_\star >0$ such that whenever $\beta>\beta_\star$, we have
$
\mathds{P} \big( \lim_{t\rightarrow \infty} \max_{i\in \mathcal{V}} | s_i(t)| = \infty\big)=1
$
for almost all initial states $s(0)$ (under the standard Lebesgue measure).
\end{thm}

Theorem \ref{thmsr3} shows that under appropriate conditions, $\max_{i\in \mathcal{V}} | s_i(t)|$ diverges almost surely if the negative updating parameter $\beta$ is sufficiently large. Actually, one may even prove that when $\max_{i\in \mathcal{V}} | s_i(t)|$ grows large when $t\to\infty$, the state of  any node diverges. This result is referred to as the {\it no-survivor} property, and is formally stated in the following proposition.

\begin{prop}\label{prop1}
Assume that A1, A2 and A6 hold, and that  for any $t\ge 0$, $b_t\equiv b$ and $d_t\equiv d$ for some $b,d \in (0,1)$. Fix the initial state $s(0)$. Then under the state-reversion  model, we have
$$
\mathds{P} \Big( \limsup_{t\rightarrow \infty} | s_i(t)| = \infty, i\in\mathcal{V} \ \Big| \limsup_{t\rightarrow \infty} \max_{i\in \mathcal{V}} | s_i(t)| = \infty\Big)=1.
$$
\end{prop}

In all above results, it can be seen from their proofs that extensions to time-varying  $\{b_t\}_{\geq 0}$ and $\{d_t\}_{\geq 0}$ are straightforward under mild assumptions. The resulting expressions are however more involved. We omit those discussions  here to simplify  the presentation.
\subsection{Supporting Lemmas}

Before proving the presented results, we first provide a few lemmas that will prove instrumental. For any $t\ge 0$, we define $M(t)=\max_{i\in \mathcal{V}} |s_i(t)|$.

\begin{lem}\label{lem1}
 Suppose $\alpha+\beta \leq {1}/{(n-1)}$. Then $M(t+1)\le M(t)$.
\end{lem}
\noindent{\it Proof.} Define $Y(t)=\alpha B_t|N_i^+(t)| +\beta  D_t|N_i^-(t)|$. Observe that  $|N_i^+(t)|+|N_i^-(t)|\leq n-1$, and hence $Y(t)\in [0,1]$ as long as $\alpha+\beta \leq {1}/{(n-1)}$. Now for any $i\in {\cal V}$,
\begin{align}
|s_i(t+1)|&\leq \Big( \big| 1-Y(t) \big| +Y(t) \Big)  \max_{j\in \mathcal{V}} |s_j(t)|\nonumber\\
&= \max_{j\in \mathcal{V}} |s_j(t)|, \nonumber
\end{align}
which completes the proof. \hfill$\blacksquare$

\begin{lem}\label{lem2}
Assume that $\alpha+\beta \leq {1}/{(n-1)}$. Let $i\in\mathcal{V}$ and assume that $|s_i(t)|\leq  \zeta _0 M(t)$ for some $0<\zeta_0<1$. Then
$$
 |s_i(t+ k)|\leq  \big(1-  (1-\zeta_0)\gamma_\ast^k \big) M(t), \quad k=0,1,\dots
$$
 where $\gamma_\ast=1- (\alpha+\beta)(n-1)$.
\end{lem}
\noindent{\it Proof.} Let $Y(t)$ be as defined in Lemma 1. We have:
\begin{align}
&|s_i(t+1)|\leq \Big( 1-Y(t) \Big) |s_i(t)|  + Y(t) M(t)\nonumber\\
&\leq \Big( 1-Y(t) \Big) \xi_0M(t)  + Y(t) M(t)\nonumber\\
&\leq  \Big(1-(\alpha+\beta)(n-1) \Big) \zeta_0 M(t)  +(\alpha+\beta)(n-1) M(t) \nonumber\\
&=\big(1-  (1-\zeta_0)\gamma_\ast \big) M(t). \nonumber
\end{align}
The lemma is then obtained by applying a simple induction argument. \hfill$\blacksquare$

\begin{lem}\label{lem3}
Assume that $\alpha+\beta \leq {1}/{(n-1)}$. Let $i\in\mathcal{V}$ and assume that $|s_i(t)|\leq  \zeta _0 M(t)$ for some $0<\zeta_0<1$. Let $(i,j)\in\mathcal{E}_t$. Then conditioned on
 $B_t=1$ if $(i,j) \in \mathcal{E}_t^+$, $D_t=1$ if $(i,j) \in \mathcal{E}_t^-$, we have
$$
|s_j(t+ 1)| \leq  \big(1-  (1-\zeta_0)\min\{\alpha,\beta\}\big) M(t).
$$
\end{lem}
\noindent{\it Proof.} Based on the update rule it can be easily seen that:
\begin{align}
|s_j(t+1)|&\leq  \min\{\alpha,\beta\}  |s_i(t)|  + \big(1- \min\{\alpha,\beta\} \big) M(t).\nonumber
\end{align}
Plugging in $|s_i(t)|\leq  \zeta _0 M(t)$, one gets the desired result. \hfill$\blacksquare$

Note that if the conditions in Lemmas \ref{lem2} and \ref{lem3} are replaced by $|s_i(t)|<  \zeta _0 M(t)$, then we have the same conclusions but with strict inequalities. Moreover, in view of Lemma \ref{lem1}, the following limit is well defined: $M_\ast= \lim_{t\rightarrow \infty} M(t)$.

\begin{lem}\label{lem4}
Assume that A1 and A2 hold, $\alpha,\beta>0$, and  $\alpha+\beta  \leq {1}/{(n-1)}$. Further assume that for any $t\ge 0$, $b_t\equiv b$ and $d_t\equiv d$ for some $b,d \in (0,1)$. Then for any initial state $s(0)$, we have $\mathds{P} \big( \lim_{t\rightarrow \infty} |s_i(t)|=M_\ast, \forall i\in\mathcal{V}\big)=1$.
\end{lem}

\noindent{\it Proof.} We prove this lemma using sample path arguments by contradiction. Let us assume that:

 \noindent {H1.} There exist $i_0\in\mathcal{V}$ and  $\delta, q_\ast\in (0,1)$ such that
$
\mathds{P} \big( \liminf_{t\rightarrow \infty} |s_{i_0}(t)|<\delta M_\ast\big)\geq q_\ast.
$

Let $\epsilon >0$. Define
$$
T(\epsilon):=\inf_{k\geq 0} \big\{ M(t)\leq (1+\epsilon) M_\ast, \forall t\geq k \big\}
$$
and
$$
T^\ast:=\inf_{t\geq T} \big\{s_{i_0}(t) <\delta M_\ast \big\}.
$$
Note that $T(\epsilon)$ is a stopping time, and the monotonicity of $M(t)$ guarantees that $T$ is bounded almost surely~\cite{durr}. Moreover, $T^\ast$ is also a stopping time, and it is bounded with probability at least $q_\ast$ in view of H1. Next, we use Lemmas \ref{lem2} and \ref{lem3} to get a contradiction. Applying Lemma~\ref{lem2},  conditioned on $\{T^\ast <\infty\}$, we have that for all $k=0,1,\dots$:
\begin{align}
\big|s_{i_0}(T^\ast+k)\big| <\big(1- (1-\delta)\gamma_\ast^k \big)  M_\ast(1+\epsilon).
\end{align}

Now  consider the time interval $[T^\ast, T^\ast+K-1]$.  The independence of  $\{{B}_t\}_{t\ge 0}$, $\{{D}_t\}_{t\ge 0}$, and $\{G_t\}_{t\ge 0}$  guarantee that
 $({G}_{T^\ast}, {B}_{T^\ast}, {D}_{T^\ast}), ( {G}_{T^\ast+1}, {B}_{T^\ast+1}, {D}_{T^\ast+1}),\dots$ are independent random variables, and they are independent of $\mathcal{F}_{{T^\ast}-1}$ (cf. Theorem 4.1.3 in \cite{durr}). From their definitions we also know that $({B}_{T^\ast}, {D}_{T^\ast})$,  $({B}_{T^\ast+1}, {D}_{T^\ast+1})$, $\dots$ are i.i.d. with the same distribution as $({B}_{0}, {D}_{0})$, and Assumption A2 guarantees that  $\mathcal{G}\big({[T^\ast,T^\ast+K-1]}\big)=\big(\mathcal{V}, \bigcup_{\tau \in [T^\ast,T^\ast+K-1] } \mathcal{E}_\tau\big)$ is strongly connected. Therefore, there exists a node $i_1\neq i_0$ and $\tau_1\leq K$ such that $(i_0,i_1) \in \mathcal{E}_{T^\ast+\tau_1}$ (note that $i_1$ and $\tau_1$ are random variables, but they are independent with $\mathcal{F}_{{T^\ast}-1}$ since $T^\ast$ is a stopping time). Hence we can apply Lemma \ref{lem3} and conclude that
\begin{align}
\big|s_{i_1}(T^\ast +\tau_1)\big| < \big(1-(1-\delta)\gamma_\ast^{\tau_1}\min\{\alpha,\beta\}\big)M_\ast(1+\epsilon) \nonumber
\end{align}
with a probability at least $p\min\{b,d\}$. Again by Lemma \ref{lem2} we have that for all $k=K,K+1,\dots$,
\begin{align}
\big|s_{i_1}(T^\ast +k)\big| < \big(1-(1-\delta)\gamma_\ast^{k}\min\{\alpha,\beta\}\big)M_\ast(1+\epsilon). \nonumber
\end{align}

We can repeat the same argument over time intervals $[T^\ast+K,T^\ast+2K-1],\dots,[T^\ast+(n-2)K,T^\ast+(n-1)K-1]$, and find $i_2,\dots,i_{n-1}$ such that $\mathcal{V}=\{i_0,\dots,i_{n-1}\}$ and bound the absolute values of their states. Finally, we get:
\begin{align}\label{1}
&\mathds{P} \big(M(T^\ast +(n-1)K) <  \big[1-\gamma_\ast^{(n-1)K}  (\min\{\alpha,\beta\})^{n-2} \nonumber\\
& \times(1-\delta)\big]  M_\ast(1+\epsilon) \big| T^\ast <\infty\big)\geq \big (p_\ast \min\{b,d\}\big)^{n-1}.
\end{align}
Now select $\epsilon$ sufficiently small so that $ \theta:=\big(1-(1-\delta)\gamma_\ast^{(n-1)K}(\min\{\alpha,\beta\})^{n-2}\big)(1+\epsilon)<1$. Using the monotonicity of $M(t)$ established in Lemma \ref{lem1}, we deduce from (\ref{1}):
\begin{align}
\mathds{P} \big(M_\ast < \theta M_\ast \big| T^\ast<\infty\big)\geq \big (p_\ast\min\{b_\ast,d_\ast\}\big)^{n-1}, \nonumber
\end{align}
which is impossible and hence, H1 is not true. We have proved that:
$$
\mathds{P} \big( \liminf_{t\rightarrow \infty} |s_{i}(t)|= M_\ast, \forall i\in\mathcal{V}\big)=1.
$$
The claim then follows easily from Lemma \ref{lem1}. \hfill$\blacksquare$

\begin{lem}\label{lem5}
Let $\alpha<(2n)^{-1}$ and $\beta>16n^{n+1}$. Then $M(t+1)\geq (2n)^{-1} M(t) $.
\end{lem}
\noindent{\it Proof.} Let us first assume that $D_t=0$. Let $i\in\mathcal{V}$ such that $|s_i(t)|=M(t)$. Then with $\alpha<(2n)^{-1}$, we have
\begin{align}
&M(t+1)\geq|s_i(t+1)|\nonumber\\
&\geq \big|1- \alpha B_t |N_i^{+}(t)| \big| \cdot|s_i(t)|-\alpha B_t |N_i^{+}(t)|\cdot M(t)\nonumber\\
&\geq \big|1- 2\alpha B_t |N_i^{+}(t)| \big| \cdot M(t) \nonumber\\
&\geq {1\over n} M(t) \geq (2n)^{-1}M(t). \nonumber
\end{align}

Now assume that $D_t=1$. We first prove the following claim.

\noindent
{\it Claim.} Consider $i_1$ such that $s_{i_1}(t) \in\big[(1-Z_2  )M(t), (1-Z_1  )M(t)\big]$ with $0\leq Z_1  <Z_2  <nZ_2  <1/4$ and $\beta Z_2  \geq 2$. Then $\mathcal{H}_1\bigcup \mathcal{H}_2$ is a sure event, where
$$
\mathcal{H}_1=\big\{M(t+1)\geq M(t)/4\big\}
$$
and
$$
\mathcal{H}_2=\big\{\exists i_2 : s_{i_2} (t)\in \big(-(1-Z_2  )M(t),-(1-nZ_2  )M(t)\big)\big \}.
$$
To prove this claim, we distinguish two cases:\\
(i) Assume that there exists $j_\ast\in\mathcal{V}$ such that $j_\ast\in N^-_{i_1}(t)$ and $s_{j_\ast}(t)\in \big[-(1-nZ_2  ) M(t), M(t)\big]$.
Then $s_{i_1}(t)+s_{j_\ast}(t)\geq (n-1)Z_2  M(t)\geq0$ and $s_{i_1}(t)+s_{j}(t)\geq -Z_2  M(t)$ for all $j\in \mathcal{V}\setminus\{i_1,j_\ast\}$. Thus, taking out the term $s_{i_1}(t)+s_{j_\ast}(t)$ in $h_{j_\ast}^-(t)$ from (\ref{9}), some simple algebra leads to
\begin{align}\label{6}
&M(t+1)\geq |s_{j_\ast}(t+1)| \nonumber\\
 &\geq  \beta\big|s_{i_1}(t)+s_{j_\ast}(t)\big|-M(t)-\alpha(n-1)M(t)\nonumber\\
 &\ \ -\beta (n-2)Z_2 M(t) \nonumber\\
&\geq \big| \beta Z_2  -1 - ({n-1})({2n})^{-1}\big|\cdot M(t)\nonumber\\
&\geq \frac{1}{2} M(t).
\end{align}

(ii) Assume that $s_{j}(t)\in[-M(t),-(1-Z_2)(M(t))]$ for all  $j\in N^-_{i_1}(t)$. Then $s_{i_1}(t)+s_{j}(t)\leq - Z_1 M(t)\leq 0$ for all $j\in N^-_{i_1}(t)$, which implies that $h_{i_1}^{-}(t)\geq 0$. Observing that $s_{i_1}(t)\geq 0$, we obtain
\begin{align}\label{7}
&M(t+1)\geq |s_{i_1}(t+1)| \nonumber\\
 &\geq |s_{i_1} (t)|- \alpha(n-1)M(t)\nonumber\\
&\geq \big| 1- Z_2   - ({n-1})({2n})^{-1}\big|\cdot M(t)\nonumber\\
&\geq \frac{1}{4} M(t).
\end{align}

From (\ref{6}) and (\ref{7}), we deduce that if $\mathcal{H}_2$ does not hold, then $\mathcal{H}_1$ is true, which proves the claim.

Finally,  we complete the proof of the lemma using the claim we just established. Take $\epsilon= 8^{-1} n^{-n-1}$ and $\beta=16 n^{n+1}$. We proceeds in steps.\\
(1) Let $m_1 \in \mathcal{V}$ with $|s_{m_1}(t)|=M(t)$. Without loss of generality, by symmetry we can assume that $s_{m_1}(t)=M(t)$. Applying the claim with $Z_1=0$ and $Z_2=\epsilon$, we deduce that either the lemma holds or there is another node $m_2 \in \mathcal{V}$ such that $s_{m_2}(t)\in \big(-(1- \epsilon  )M(t),-(1-n \epsilon )M(t)\big)$.\\
(2) If in the first step, we could not conclude that the lemma holds, we can apply the claim to $m_2$ (observe that the claim we established is also valid when all states $s_i(t)$ are replaced by $-s_i(t)$). We then obtain that either the lemma holds, or there is a node $m_3$ such that  $s_{m_3}(t)\in \big( (1-n^2 \epsilon )M(t),(1- n\epsilon)M(t)\big)$.\\
The argument can be repeated for $m_3,\dots$ applying the claim adapting the value of $\epsilon$ and $\beta$. Since the number of nodes is bounded, the above repeated procedure necessarily ends, so the lemma holds. \hfill$\blacksquare$

\subsection{Proofs of the Main Results}

\subsubsection*{Proof of Theorem \ref{thmsr1}} From Lemma \ref{lem4}, we know that for any $i\in\mathcal{V}$, one of the following events happens almost surely:
$\big\{ \lim_{t\rightarrow \infty} s_{i}(t)= M_\ast\}$; $\big\{ \lim_{t\rightarrow \infty} s_{i}(t)=-M_\ast\}$; $\big\{ \liminf_{t\rightarrow \infty} s_{i}(t)= -M_\ast$ and $ \limsup_{t\rightarrow \infty} s_{i}(t)= M_\ast\}$. Therefore, we just need to rule out the last case. We actually prove that:
 $
 \mathds{P}\big(M_\ast>0, \liminf_{t\rightarrow \infty} s_{i}(t)= -M_\ast, \limsup_{t\rightarrow \infty} s_{i}(t)= M_\ast, \lim_{t\rightarrow \infty} |s_{i}(t)|= M_\ast\big)=0.
 $

Let $\epsilon>0 $ and define $T_1(\epsilon):=\inf\big\{ k : M(k)\leq M_\ast(1+\epsilon)\big\}$. Using a similar recursive argument as that used in  the proof of Lemma \ref{lem2}, we get: for all $k=0,1,\dots$ and $t\geq T_1$,
\begin{align}\label{2}
s_i(t+k)&\leq \gamma_\ast^k s_i(t)+ (1-\gamma_\ast^k)M_\ast(1+\epsilon ).
\end{align}

Let $M_\ast >0$. Assume that $\liminf_{t\rightarrow \infty} s_{i}(t)= -M_\ast$. Then for the given $\epsilon$, we can find an infinite sequence $T_1(\epsilon)<t_1<t_2<\dots$ such that $s_{i}(t_m) \leq -M_\ast(1-\epsilon)$. Now, if $\limsup_{t\rightarrow \infty} s_{i}(t)= M_\ast$, for any $t_m$, we can find $\bar{t}_m>t_m$ with $s_{i}(\bar{t}_m)\geq  M_\ast(1-\epsilon)$. Then based on (\ref{2}), there must be $\hat{t}_m \in[t_m,\bar{t}_m]$ such that
$
s_i(\hat{t}_m)\in \big[-\gamma_\ast M_\ast(1-\epsilon)+ (1-\gamma_\ast)M_\ast(1+\epsilon ), -\gamma_\ast^2 M_\ast(1-\epsilon)+ (1-\gamma_\ast^2)M_\ast(1+\epsilon )\big].
$
We deduce that $|s_i(\hat{t}_m)|<M_\ast\big(1+ \max\{|1-2\gamma_\ast|,|1-2\gamma_\ast^2|\}\big)/2$, $m=1,2,\dots$, when $\epsilon<\big(1- \max\{|1-2\gamma_\ast|,|1-2\gamma_\ast^2|\}\big)/2$. This contradicts  $\lim_{t\rightarrow \infty} |s_{i}(t)|= M_\ast$ since by our assumption we have $0<\gamma_\ast<1$.\hfill$\blacksquare$

\subsubsection*{Proof of Theorem \ref{thmsr2}}  In view of Theorem \ref{thmsr1}, at least one of the following sets is non-empty:
$$
\mathcal{V}_1^\ast:=\big \{i\in\mathcal{V}: \lim_{t \rightarrow \infty} s_i(t)=-M_\ast\big \},
 $$
 and
 $$\mathcal{V}_2^\ast:=\big \{i\in\mathcal{V}: \lim_{t \rightarrow \infty} s_i(t)=M_\ast\big \}.
$$
Without loss of generality, we assume $\mathcal{V}_1^\ast \neq \emptyset$ and $\mathds{P}(M_\ast>0)>0$.

\noindent {(i).} Applying the same sample-path analysis as in the proof of Theorem \ref{thmsr1}, one can easily show that the arcs among nodes in $\mathcal{V}_1^\ast$ are necessarily positive since each negative link appears for infinite time slots. Now the total graph $\mathcal{G}_\ast$ is strongly balanced with nonempty $\mathcal{V}_1$ and $\mathcal{V}_2$, and hence ${\cal V}_1^\ast$ is for example included in ${\cal V}_1$, which in turns implies that ${\cal V}_2^\ast \neq \emptyset$. Again there are only positive arcs among  nodes of ${\cal V}_2^\ast$. We simply deduce that $\{\mathcal{V}_1,\mathcal{V}_2\}= \{ \mathcal{V}_1^\ast,\mathcal{V}_2^\ast\}$.

\noindent {(ii).} Since $\mathds{P}(M_\ast >0)>0$, we have $\mathcal{V}_1^\ast\cap \mathcal{V}_2^\ast=\emptyset$.  Again arcs between nodes in the same set from $\mathcal{V}_i^\ast, i=1,2$ are necessarily positive. However there is at least one negative link in $\mathcal{G}_\ast$ by our assumption, which can only be an arc between $\mathcal{V}_1^\ast$ and $\mathcal{V}_2^\ast$.   Thus both $\mathcal{V}_1^\ast$ and $\mathcal{V}_2^\ast$ are nonempty, which implies that $\mathcal{G}_\ast$ must be strongly balanced.  This contradicts our standing assumption and the proof is complete.  \hfill$\blacksquare$

\subsubsection*{Proof of Theorem \ref{thmsr3}} Let $\beta>16n^{n+1}$ so the conditions of Lemma \ref{lem5} hold. Let us fix $t\geq 0$ and assume that $|s_{i_0}(t)|=M(t)$ for some $i_0 \in \mathcal{V}$. By symmetry, we can also assume without loss of generality that $s_{i_0}(t)=M(t)$. Let $i_\ast \in \mathcal{V}\setminus\{i_0\}$. Under Assumptions A4 and A6, we prove the following claim.

\noindent{\it Claim.} There is an integer $N_0\geq 1$ and $q_0>0$ such that
$$
\mathds{P}\Big(s_{i_0}(t+N_0K)=M(t),s_{i_\ast}(t+N_0K) \geq {M(t)}/{2}\Big) \geq q_0.
$$

In view of the connectivity condition A4 and the arc independence condition A6, the event $\{s_{i_\ast}(t+N_0K-1) \geq {M(t)}/{2}\}$ given  $s_{i_0}(t)=M(t)$ can be easily constructed by selecting a proper sequence of positive arcs for time slots $t,t+1,\dots,t+N_0K-1$, and by imposing that $B_\tau=1,D_\tau=0,$ $\tau=t,t+1,\dots,t+N_0K-1$. Here $N_0$ and $q_0$ depend on $\alpha,b_\ast,d_\ast,p_\ast,p^\ast,n$ but do not rely on $\beta$. The analysis follows arguments to analyze basic consensus algorithms, and we omit the details.

In addition, in view of Assumption A5, we can select a node $i_\ast \neq i_0$ satisfying  $(i_\ast,i_0) \in \bigcup_{\tau \in [t+N_0K,t+(N_0+1)K-1] } \mathcal{E}^-_\tau$. It then follows that
\begin{align}
&\mathds{P}\Big(|s_{i_0}(t+(N_0+1)K)|\geq \big(\frac{3}{2}\beta-1-\frac{n-1}{2n}\big) M(t) \Big)\nonumber\\
&\geq  \mathds{P}\Big( s_{i_0}(t+N_0K)=M(t),s_{i_\ast}(t+N_0K) \geq {M(t)}/{2}\Big)\nonumber\\
& \times  \mathds{P}\Big(  \exists \tau\in[t+N_0K,t+(N_0+1)K-1]\ \mbox{s.t.}\ (i_\ast,i_0)\in E^-_\tau \Big)\nonumber\\
& \times \mathds{P}(D_\tau=1) \nonumber\\
& \times \mathds{P}\Big(B_m=D_m=0,m\neq \tau\in[t+N_0K,t+(N_0+1)K-1]\Big)\nonumber\\
&\geq \vartheta_0, \nonumber
\end{align}
where $\vartheta_0=q_0 p_\ast d \big((1-d)(1-b)\big)^{K-1}$. This implies
\begin{align}\label{10}
\mathds{P}\big(M(t+N_0K)\geq  3(\beta-1) M(t)/2  \big) \geq \vartheta_0.
\end{align}

Now assume that $M(0)>0$ so that $U(m)=\log(M(mN_0K))$ for $m\ge 0$ is well defined. Note that from Lemma \ref{lem5} and (\ref{10}), we have:
$$
\mathds{E}\{U(m+1)-U(m)\} \ge -N_0K\log ({2n}) +  \vartheta_0 \log \big( {3}(\beta-1) /2\big).
$$
For $\beta$ large enough, the r.h.s. in the above inequality is strictly positive. We can then easily conclude, using classical arguments in random walks that the process $U(m)$ has a strictly positive drift, from which it can be deduced that $\mathds{P}\big( \liminf_{m\rightarrow \infty} M\big(mN_0K\big)=\infty \big)=1$ (for $\beta$ large enough). Using Lemma \ref{lem5}, one can easily conclude the desired theorem. \hfill$\blacksquare$

%We can now complete the proof with the help of the strong law of large numbers. Suppose $M(0)>0$ and define
%$$
%U(m)= \frac{M\big((m+1)N_0K\big)}{M\big(mN_0K\big)},\ m=0,1,\dots.
%$$
%With Lemma \ref{lem5} and (\ref{10}), we have
%$$
%\mathds{E}\big\{\log U(m)\big\} \geq N_0K\log \Big(\frac{1}{2n}\Big) +  \vartheta_0 \log \Big( %\frac{3}{2}(\beta-1)\Big):= O(\beta).
%$$

%We can always select $\beta$ sufficiently large so that $O(\beta)>0$. Since (\ref{10}) corresponds directly to $\{G_\tau,B_\tau,D_\tau\}_{t}^{t+N_0K-1}$, while $\{G_t,B_t,D_t\}_0^\infty$ is independent. Thus, we can assume without loss of generality that $U(m),m=0,1,\dots$ are independent (otherwise we can redefine a random variable $\{G_t,B_t,D_t\}_{mK_0}^{(m+1)N_0K-1}$ from associated with each $U(m)$.)

%Lemma \ref{lem5} and the linearity of the algorithm ensure that we can find a constant %
%$U_\ast>0$ such that $|\log U(m)(\omega)|\leq U_\ast$ for all $\omega$. Thus, %$\mathds{V}\big\{\log U(m)\big\}, m=0, 1,\dots$ are bounded uniformly in $m$. Applying %Kolmogorov's strong law of large numbers under Kolmogorov criterion (see \cite{feller}) lead to %\begin{align}
 %\mathds{P}\Big( \lim_{t\rightarrow \infty} \frac{1}{t}\sum_{s=0}^t\Big(\log U(m) %-\mathds{E}\big\{ \log U(m)\big\}\Big) = 0\Big)=1,
%\end{align}

\subsubsection*{Proof of Proposition \ref{prop1}} Assume that  for some $q_\ast>0$ we have $\mathds{P}(\limsup_{t\rightarrow \infty} \max_{i\in \mathcal{V}} | s_i(t)| = \infty)\geq q_\ast$. There must be a node $i_0$ satisfying $\mathds{P}(\limsup_{t\rightarrow \infty}  | s_{i_0}(t)| = \infty)\geq q_\ast/n$. Let $C_0>0$, and define
$
T_1^\star:=\inf_{t}\big\{|s_{i_0}(t)| \geq C_0\big\}.
$
$T_1^\star$ is a stopping time. Let $Y_0>0$ be an integer. We can further recursively define $T_2^\star,\dots,T_m^\star,\dots$ by
$$
T_{m+1}^\star:=\inf_{t\geq T_m^\ast+Y_0}\big\{|s_{i_0}(t)| \geq C_0\big\}.
$$
Based on  Theorem 4.1.3 in \cite{durr}, each $T_{m}^\star$ is a stopping time for all $m\geq 0$ and $({G}_{T_1^\star}, {B}_{T_1^\star}, {D}_{T_1^\star})$, $\dots$, $( {G}_{T_1^\star+Y_0-1}, {B}_{T_1^\star+Y_0-1}, {D}_{T_1^\star+Y_0-1})$; $({G}_{T_2^\star}, {B}_{T_2^\star}, {D}_{T_2^\star}),\dots,( {G}_{T_2^\star+Y_0-1}, {B}_{T_2^\star+Y_0-1}, {D}_{T_2^\star+Y_0-1});\dots$ are independent random variables that are also independent of $\mathcal{F}_{{T_1^\star}-1}$. In addition, we have $\mathds{P}( T_m^\star<\infty, m=1,2,\dots)\geq q_\ast/n$. Under Assumption A5, $\mathcal{G}\big({[T_1^\star,T_1^\star+K-1]}\big)$
%$=\big(\mathcal{V}, \bigcup_{\tau \in [T_1^\star,T_1^\star+K-1] } \mathcal{E}_\tau\big)$
being  strongly connected is a sure event. As a result, there exists another node $i_1\in\mathcal{V}\setminus {i_0}$ and $\tau_0\in [T_1^\star,T_1^\star+K-1]$ such that $(i_0,i_1)\in \mathcal{E}_{\tau_0}$. Assume that $s_{i_0} (\tau_0)=s_{i_0}(T_1^\ast)$. We treat two cases: $\sigma_{i_0i_1}=-$ and $\sigma_{i_0i_1}=+$.
 \begin{itemize}
 \item[(i)] $\sigma_{i_0i_1}=-$.
 \begin{itemize}
\item If $\beta=1$, then $|\beta s_{i_0} (\tau_0) + (1-\beta) s_{i_1} (\tau_0)|= |\beta s_{i_0} (\tau_0)|=|s_{i_0}(T_1^\ast)|  \geq {C_0}$;

\item  If $\beta\neq 1$ and $|s_{i_1} (\tau_0)|< {\beta C_0}/{(2|1-\beta|)}$, then $|\beta s_{i_0} (\tau_0) + (1-\beta) s_{i_1} (\tau_0)|\geq \beta C_0 -(1-\beta)|s_{i_1}(\tau_0)| \geq \beta C_0/2$.
\end{itemize}
\item[(ii)] $\sigma_{i_0i_1}=+$.
\begin{itemize}
\item If $\alpha=1$, then $|\alpha s_{i_0} (\tau_0) + (1-\alpha) s_{i_1} (\tau_0)|= C_0$.

\item  If $\alpha\neq 1$ and $|s_{i_1} (\tau_0)|<{\alpha C_0}/{(2|1-\alpha|)}$, then $|\alpha s_{i_0} (\tau_0) + (1-\alpha) s_{i_1} (\tau_0)|\geq \alpha C_0/2$.
    \end{itemize}
\end{itemize}
Now $s_{i_1}(\tau_0+1)=-\beta s_{i_0} (\tau_0) + (1-\beta) s_{i_1} (\tau_0)$ when $i_0$ is the unique node in  $N_{i_1}^-({\tau_0})$ and $D_{\tau_0}=1$. Also observe that
$s_{i_1}(\tau_0+1)=\alpha s_{i_0} (\tau_0) + (1-\alpha) s_{i_1} (\tau_0)$ when $i_0$ is the unique node in  $N_{i_1}^+({\tau_0})$ and $B_{\tau_0}=1$. Independence  ensures that $({B}_{T_1^\star}, {D}_{T_1^\star})$, $\dots$, $({B}_{T_1^\star+Y_0-1}, {D}_{T_1^\star+Y_0-1})$ have the same distribution as $({B}_{0}, {D}_{0})$. We can therefore simply bound the probabilities of the above events and establish
\begin{align}
\mathds{P}\big(\exists i_1\in \mathcal{V}\setminus\{i_0\} : |s_{i_1}(T_1^\star+K)|\geq \phi C_0\big)&\geq \chi_0, \nonumber
\end{align}
where $\chi_0=\big((1-b)(1-d)\big)^{2K-1} \min\{b,d\} p_\ast(1-p^\ast)^{n-2}$ and $\phi=\min \big\{ [ {\alpha }/{(2|1-\alpha|)}], \alpha/2, [{\beta }/{(2|1-\beta|)}], \beta/2,1\big\}$ (we use $[\cdot]$ to indicate that the corresponding term is taken into account in the $\min$ only if it is well defined). Repeating the analysis on $T_2^\ast,\dots$ we obtain
\begin{align}
\mathds{P}\big(\exists i_m\in \mathcal{V}\setminus\{i_0\} : |s_{i_m}(T_m^\star+K)|\geq \phi C_0\big)\geq \chi_0. \nonumber
\end{align}
Since we have a finite number of nodes,  independence
%of  $({G}_{T_1^\star}, {B}_{T_1^\star}, {D}_{T_1^\star})$, $\dots$, $( {G}_{T_1^\star+Y_0-1}, {B}_{T_1^\star+Y_0-1}, {D}_{T_1^\star+Y_0-1})$, $({G}_{T_2^\star}, {B}_{T_2^\star}, {D}_{T_2^\star}),\dots,( {G}_{T_2^\star+Y_0-1}, {B}_{T_2^\star+Y_0-1}, {D}_{T_2^\star+Y_0-1}),\dots$
allows us to invoke the second Borel-Cantelli Lemma (cf. Theorem 2.3.6 in \cite{durr}) and conclude that
\begin{align}\label{16}
&\mathds{P}\big(\exists \ {\rm (deterministic)}\ i_1\in\mathcal{V}\setminus\{i_0\} : \nonumber\\
  &\ \ \limsup_{t\rightarrow \infty}|s_{i_1}(t)|\geq \phi C_0 \big|T_m^\ast <\infty,m=1,\dots\big)=1.
\end{align}

Note that $C_0$ can be chosen arbitrarily, and hence (\ref{16}) implies that there exists $i_1\in\mathcal{V}\setminus\{i_0\}$ such that
\begin{align}\label{eq:cc}
\mathds{P}\big(\limsup_{t\rightarrow \infty}|s_{i_1}(t)|=\infty \big|\limsup_{t\rightarrow \infty} \max_{i\in \mathcal{V}} | s_i(t)| = \infty\big)=1.
\end{align}
We can apply the same argument recursively, to show that (\ref{eq:cc}) holds for any node $i_1$ in the network.  \hfill$\blacksquare$

\section{The Relative-State-Reversion  Model}

In this section, we investigate the system dynamics under the relative-state-reversion model, and provide, as  for the state-reversion model, conditions for convergence and divergence of the node states.

\subsection{Main Results}

The following theorem provides general conditions for convergence and divergence.

\begin{thm}\label{thmrsr1}  Assume that for any $t\ge 0$, $\mathcal{G}_t\equiv \mathcal{G}   $ for some digraph $\mathcal{G}$, and that each positive cluster of $\mathcal{G}  $ admits a spanning tree in $\mathcal{G}  ^+$. Further assume that A1 holds and that $\alpha \in (0, (n-1)^{-1})$. Under the relative-state-reversion model, we have:\\
(i) If $\sum_{t=0}^\infty d_t <\infty$, then $\mathds{P} \big( \lim_{t\rightarrow \infty}s_i(t) \ {\rm exits}\big)=1$ for all node $i\in {\cal V}$ and all initial states $s(0)$;\\
(ii) If $\sum_{t=0}^\infty d_t =\infty$, then there is an infinite number of initial states $s(0)$ such that, as long as $\beta>0$ and $\mathcal{G}  $ has at least two positive clusters,
\begin{align}\label{18}
\mathds{P} \big( \lim_{t\rightarrow \infty}\max_{i,j\in \mathcal{V}} | s_i(t)-s_j(t)|=\infty\big)=1.
\end{align}
\end{thm}

The first part of the above theorem indicates that when the environment is frozen, and when positive clusters are properly  connected, then irrespective of the mean of the positive attentions $\{b_t\}_0^\infty$, the system states converge if the attention each node puts in her negative neighbors  decays sufficiently fast over time. The second part of the theorem states that when this attention does not decay, divergence is typically observed. We can easily  build examples showing that essentially, the conditions in Theorem \ref{thmrsr1} cannot be relaxed.

Next, we provide a sufficient condition for weak consensus (meaning that the distances among the node states converge to zero almost surely). It is based on the following assumption.

\noindent {\bf A7.}   There is an integer $K\geq 1$ such that the union graph $\mathcal{G}^+\big([t,t+K]\big)=\big(\mathcal{V}_i, \bigcup_{\tau \in [t,t+K-1] } \mathcal{E}_\tau^+\big)$ has a spanning tree for all $t\geq0$.

\begin{thm}\label{thmrsr2}  Assume that  A1 and A7 hold and that $\alpha \in (0,(n-1)^{-1})$. Denote $K_0=(2n-3)K$ and $\rho_\ast=\min\{\alpha, 1-(n-1)\alpha\}$. Define
$X_m=\frac{p_\ast^{n-1} \rho_\ast^{K_0}}{2}  \prod_{t=mK_0}^{(m+1)K_0-1}\big(b_t(1-d_t)\big)$, and $Y_m=\big(1+2\beta(n-1) \big)^{K_0}\big(1- \prod_{t=mK_0}^{(m+1)K_0-1}(1-d_t)\big) $.  Then under the relative-state-reversion model, if $0\leq  X_m-Y_m \leq 1$ for all $m\geq 0$ and  $\sum_{m=0}^\infty (X_m-Y_m)=\infty$, we have
$\mathds{P} \big( \limsup_{t\rightarrow \infty} \max_{i,j\in \mathcal{V}} | s_i(t)-s_j(t)| = 0\big)=1$
for  all initial states.
\end{thm}

A direct consequence of Theorem \ref{thmrsr2} is that  if  $b_t\equiv b  $ and $d_t\equiv d$ with $b  ,d \in (0,1)$ and $\beta>0$, there exists $d_\star >0$ such that whenever $d<d_\star$, weak consensus is achieved almost surely. Observe that   weak consensus  does not necessarily guarantee the convergence of the state of each node. In fact, simple examples can be constructed with arbitrarily small $\beta$ such that under the relative-state-reversion model, the state of each node grows arbitrarily large while weak consensus still holds. This contrasts  the result for the state-reversion model:  the condition  $\alpha+\beta<(n-1)^{-1}$  of Theorem \ref{thmsr1} prevents the state of individual nodes to diverge.
%Therefore, the negative {\it weight} $\beta$,  and {\it  faith} $\{d_t\}_{t\geq 0}$ play quite different roles in the state-reversion  and the relative-state-reversion models.

Next we provide conditions under which the maximal gap between the states of two nodes grows large almost surely, and establish a no-survivor property.

\noindent {\bf A8.}  There is an integer $K\geq 1$ such that the union graph $\mathcal{G}^-\big({[t,t+K]}\big)=\big(\mathcal{V}, \bigcup_{\tau \in [t,t+K-1] } \mathcal{E}_\tau^-\big)$ is weakly connected for all $t\geq0$.

\begin{thm}\label{thmrsr3} Assume that A1, A6, and A8 hold and that $\alpha \in [0,(2(n-1))^{-1})$. Let $b_t\equiv b  $ and $d_t\equiv d$ for some constants $b  ,d \in (0,1)$. Let $\beta >0$ and fix $d$. Then under the relative-state-reversion, there is $b_\star >0$ such that whenever $b  <b_\star$, we have    $
\mathds{P} \big( \lim_{t\rightarrow \infty} \max_{i,j\in \mathcal{V}} | s_i(t)-s_j(t)| = \infty\big)=1
$
for almost all initial states (under the standard Lebesgue measure).
\end{thm}

\begin{prop}\label{prop2}
Assume that  A1, A4 and A6 hold. Let $b_t\equiv b  $ and $d_t\equiv d$ for some constants $b  ,d \in (0,1)$. Let $\alpha, \beta>0$.  Fix the initial value $s(0)$. Then under the relative-state-reversion  model, we have
$
\mathds{P} \big( \limsup_{t\rightarrow \infty} | s_i(t)-s_j(t)| = \infty, i\neq j\in\mathcal{V} \big| \limsup_{t\rightarrow \infty} \max_{i,j\in \mathcal{V}} | s_i(t)-s_j(t)| = \infty\big)=1.
$
\end{prop}

%\begin{remark}
%Proposition \ref{prop2} under some additional assumptions, we can even  conclude from %(\ref{18}) that at least $n-1$ nodes achieves almost sure divergence in their states.
%\end{remark}

Finally, we investigate the clustering of states of nodes within each positive cluster.

\noindent {\bf A9.} Assume that A3 holds and let $\mathcal{V}=\bigcup_{i=1}^{{{\rm T}_{\rm p}}} \mathcal{V}_i$ be a positive-cluster partition of the total graph $\mathcal{G}_\ast$. There is an integer $K\geq 1$ such that the union graph $\mathcal{G}^+\big({[t,t+K]}\big)\big|_{\mathcal{V}_i}=\big(\mathcal{V}_i, \bigcup_{\tau \in [t,t+K-1] } \mathcal{E}_\tau^+\big|_{\mathcal{V}_i}\big)$ has a spanning tree for all $t\geq0$.

\begin{thm} \label{thmrsr4} Assume that A1, A3 and A9 hold and let $\mathcal{V}=\bigcup_{i=1}^{{{\rm T}_{\rm p}}} \mathcal{V}_i$ be a positive-cluster partition of $\mathcal{G}_\ast$. Let $\alpha \in (0,(n-1)^{-1})$. Define $J(m)=\prod_{t=mK_0}^{(m+1)K_0-1}b_t$ and $W(m)=\sum_{t=mK_0}^{(m+1)K_0-1} d_t$ with $K_0=(2n-3)K$. Further assume that $\sum_{m=0}^\infty J(m)=\infty$, $\sum_{t=0}^\infty d_t   <\infty$, and $\lim_{m \rightarrow\infty}W(m)/J(m)=0$. Then under the relative-state-reversion model, for any initial state $s(0)$, there are ${{\rm T}_{\rm p}}$ real-valued random variables, $w^\ast_1,\dots,w^\ast_{{{\rm T}_{\rm p}}}$, corresponding to each of the positive clusters, such that $\mathds{P} \big( \lim_{t\rightarrow \infty }s_i(t)= w^\ast_j,\ i\in\mathcal{V}_j,\ j=1,\dots, {{\rm T}_{\rm p}}\big)=1$.
\end{thm}

\subsection{Supporting Lemmas}

We list three martingale convergence lemmas (see e.g. \cite{polyak}), and a result that will be instrumental in the analysis of the system convergence under the relative-state-reversion model.

\begin{lem}\label{lem6} Let $\{v_t\}_{t\ge 0}$ be a sequence of non-negative random variables with $\mathds{E}\{v_0\}<\infty$. Assume that for any $t\ge 0$,
$$
\mathds{E}\{v_{t+1}|v_0,\dots,v_t\}\leq (1+\xi_t)v_t+\theta_t,
$$
where $\{\xi_t\}_{t\ge 0}$ and $\{\theta_t\}_{t\ge 0}$ are two (deterministic) sequences of non-negative numbers satisfying $\sum_{t=0}^\infty {\xi_t}<\infty$ and $\sum_{t=0}^\infty {\theta_t}<\infty$. Then $\lim_{t\to\infty} v_t = v$ a.s. for some random variable $v\geq 0$.
\end{lem}

\begin{lem}\label{lem7}
Let$\{v_t\}_{t\ge 0}$ be a sequence of non-negative random variables with $\mathds{E}\{v_0\}<\infty$. Assume that for any $t\ge 0$,
$$
\mathds{E}\{v_{t+1}|v_0,\dots,v_t\}\leq (1-\xi_t)v_t+\theta_t,
$$
where $\{\xi_t\}_{t\ge 0}$ and $\{\theta_t\}_{t\ge 0}$ are two (deterministic) sequences of non-negative numbers satisfying $\forall t\ge 0$, $0\leq \xi_t \leq 1$, $\sum_{t=0}^\infty {\xi_t}=\infty$, $\sum_{t=0}^\infty {\theta_t}<\infty$, and $\lim_{t\to\infty}{\theta_t}/{\xi_t}=0$. Then $\lim_{t\to\infty} v_t = 0$ a.s..
\end{lem}

\begin{lem}\label{lem8} (Robbins-Siegmund)
Let $\{v_t\}_{t\ge 0},\{\xi_t\}_{t\ge 0},\{\theta_t\}_{t\ge 0}$ be sequences of non-negative random variables. Assume that for any $t\ge 0$,
$$
\mathds{E}\{v_{t+1}|\mathcal{F}_t\}\leq (1+\xi_t)v_t+\theta_t,
$$
where $\mathcal{F}_t=\sigma(v_0,\dots,v_t;\xi_0,\dots,\xi_t;\theta_0,\dots,\theta_t)$. Suppose $\sum_{t=0}^\infty {\xi_t}<\infty$ and $\sum_{t=0}^\infty {\theta_t}<\infty$ almost surely. Then $\lim_{t\to\infty} v_t = 0$ a.s. for some random variable $v\geq 0$.
\end{lem}

%\begin{lem}\label{lem9}
%Let $\{\xi_t\}_0^\infty$ and $\{\theta_t\}_0^\infty$ be two (deterministic) nonnegative  %sequences satisfying $\sum_{t=0}^\infty {\xi_t}<\infty$ and $\lim_{t\rightarrow \infty %}\theta_t$ exists. Then $\sum_{t=0}^\infty {\xi_t \theta_t}<\infty$.
%\end{lem}

\begin{lem}\label{lem10}
Define $h(t):= \min_{i\in\mathcal{V}} s_i(t)$, $H(t):= \max_{i\in\mathcal{V}} s_i(t)$, and $\mathcal{H}(t):=H(t)-h(t)$.  Assume that $\alpha \in [0, (n-1)^{-1}]$ and that $\sum_{t=0}^\infty d_t<\infty$. Then  under the relative-state-reversion model, for all initial states, $h(t)$, $H(t)$, $\mathcal{H}(t)$ converges almost surely as $t$ grows large.
\end{lem}

\noindent{\it Proof.} We first prove the convergence of $\mathcal{H}(t)$. On can easily see that when $\alpha \in [0, (n-1)^{-1}]$ and $D_t=0$, $H(t+1)\leq H(t)$, $h(t+1)\geq h(t)$, and thus $\mathcal{H}(t+1)\leq \mathcal{H}(t)$. When $D_t=1$, $\mathcal{H}(t+1)\leq (2\beta (n-1)+1)\mathcal{H}(t) $. We deduce that:
\begin{align}\label{100}
\mathds{E}\big\{\mathcal{H}(t+1)|\mathcal{H}(t)\big\}\leq \big(1+ 2\beta (n-1)d_t\big)\mathcal{H}(t),
\end{align}
which, in view of Lemma \ref{lem6}, implies that $\mathcal{H}(t)\rightarrow \mathcal{H}_\ast$ almost surely for some $\mathcal{H}_\ast \geq 0$. Similarly, for $H(t)$, we have
$
\mathds{E}\big\{{H}(t+1)|{H}(t)\big)\}\leq H(t)+\zeta(t),
$
where $\zeta(t):=\big(1+ 2\beta (n-1)d_t\big)\mathcal{H}(t)$. Since $\mathcal{H}(t)$ converges a.s. and $\sum_td_t<\infty$, we deduce that $\sum_t\zeta(t)<\infty$ a.s.. Now, the first Borel-Cantelli Lemma (Theorem 2.3.1, \cite{durr}) and the fact that $H(t)\geq h(t)$ ensure that $\mathds{P}(\inf_{t\geq 0}H(t)>-\infty)=1$. In other words, $H(t)$
is almost surely lower bounded. Hence, we can still invoke Lemma \ref{lem8} to conclude that $H(t)$ converges almost surely as $t$ grows large. The convergence of $h(t)$ follows from a symmetric argument. \hfill$\blacksquare$
\subsection{Proofs of Main Results}
We now prove the various results stated  previously.

\subsubsection*{Proof of Theorem \ref{thmrsr1}} (i) We investigate two cases: $\sum_{t=0}^\infty b_t<\infty$, and $\sum_{t=0}^\infty b_t=\infty$.

a). Assume that $\sum_{t=0}^\infty b_t<\infty$. Since we also have $\sum_{t=0}^\infty d_t<\infty$, the first Borel-Cantelli Lemma guarantees that almost surely, each node revises its state for only a finite number of slots. The desired claim follows obviously.

b). Assume that $\sum_{t=0}^\infty b_t=\infty$. With $\sum_{t=0}^\infty d_t<\infty$, from the first Borel-Cantelli Lemma,
$$
K_\ast:= \inf\{k\geq 0: D_t=0, \forall t\geq k\}
$$
is almost surely bounded. We note that $K_\ast$ is not a stopping time for $\{D_t\}_{t\geq 0}$, but a stopping time for $\{B_t\}_{t\geq 0}$ by independence of $\{B_t\}_{t\geq 0}$ and $\{D_t\}_{t\geq 0}$. Hence, the second Borel-Cantelli lemma ensures that
$$
K_{m+1}:= \inf\{t\ge K_m : B_t=1\},\ m=0,1,\dots
$$
and $K_0:= \inf\{t\ge K_\ast: B_t=1\}$ are stopping times for $\{B_t\}_{t\geq 0}$. Now in view of the independence of $\{G_t\}_{t\geq 0}$, $\{B_t\}_{t\geq 0}$, and $\{D_t\}_{t\geq 0}$, we know  that $\{G_{K_m}\}_{m\geq 0}$ is an independent  process and each $G_{K_m}$ satisfies  $\mathds{P}\big( (i,j)\in E_{K_m}  \big)\geq p_\ast$ for all $(i,j)\in \mathcal{G}$ under Assumption A1.

Let $\mathcal{V}^\dag$ be a positive cluster of $\mathcal{G} $. By assumption, $\mathcal{V}^\dag$ has a spanning tree. Since $\alpha<1/(n-1)$, the above discussion shows that at times $K_m,m=0,1,\dots$, the considered relative-state-reversion model defines a standard  consensus dynamics on independent  random graphs where each arc exists with probability at least $p_\ast$ for any fixed time slot.  It has become clear from existing works on randomized consensus dynamics (as follows from connectivity-independent graphs in \cite{shiIT},  combining Theorem 1 in \cite{xiao} and Theorem 3 in \cite{jad08} for the i.i.d. case with $\mu_t \equiv \mu$, or various other implicit results in the literature \cite{it10}),  that the connectivity of $\mathcal{V}^\dag$ ensures that $\mathds{P}\big(\lim_{m \rightarrow \infty} \mathcal{H}^\dag(K_m)=0\big)=1$, where $\mathcal{H}^\dag(t)=\max_{i\in \mathcal{V}^\dag} s_i(t)-\min_{i\in \mathcal{V}^\dag} s_i(t)$. The monotonicity of $\mathcal{H}^\dag(t)$ for $t\geq K_\ast$ further ensures that $\mathds{P}\big(\lim_{t \rightarrow \infty} \mathcal{H}^\dag(t)=0\big)=1$. Finally, applying the analysis of Lemma \ref{lem10} restricted to $\mathcal{V}^\dag$, we show that $\mathds{P}\big(\lim_{t \rightarrow \infty} \max_{i\in \mathcal{V}^\dag} s_i(t)= H^\dag_\ast\big)=1$ and that $\mathds{P}\big(\lim_{t \rightarrow \infty} \min_{i\in \mathcal{V}^\dag} s_i(t)= h^\dag_\ast\big)=1$ for some  $H^\dag_\ast$ and $h^\dag_\ast$. Therefore, $H^\dag_\ast=h^\dag_\ast$ almost surely, which implies $\lim_{t \rightarrow \infty} s_i(t)= H^\dag_\ast =h^\dag_\ast$ almost surely for all $i\in \mathcal{V}^\dag$. This completes the proof.

\noindent (ii) Let $\mathcal{V}=\bigcup_{i=1}^{{\rm T}_{\rm p}} \mathcal{V}_i$ be a positive-cluster partition of $\mathcal{G}  $ with ${\rm T}_{\rm p}\geq 2$. Let $\epsilon >0$. Set $s_i(t)= j\epsilon$ for all $i\in\mathcal{V}_j$, $j=1,\dots,{\rm T}_{\rm p}$. Applying the second Borel-Cantelli lemma, the divergence condition (\ref{18}) can be simply established by investigating the evolution of $\mathcal{H}(t)$ under the assumption that $\sum_{t=0}^\infty d_t=\infty$.
\hfill$\blacksquare$

\subsubsection*{Proof of Theorem \ref{thmrsr2}} The proof relies on  Lemma \ref{lem7},  cf.,  \cite{it10} for the analysis of randomized consensus.

Consider $2n-3$ intervals $[mK,(m+1)K-1],m=0,\dots,2(n-2)$. With Assumption A7, there is a center node $v_m \in \mathcal{V}$ in each of $\mathcal{G}([mK,(m+1)K-1])$. As a result, we can find  $n-1$ center nodes  (repetitions are allowed) out of the $v_m$'s and denote them as $v_{m_1},\dots,v_{m_{n-1}}$, that satisfy  either $s_{v_{m_j}}(0) \leq (h(0)+H(0))/2$, or $s_{v_{m_j}}(0) > (h(0)+H(0))/2$, for all $j=1,\dots,n-1$. Without loss of generality, we consider the first case only.

Assume that $D_t=0$ for $t=0,\dots, 2(n-2)K-1$. The following facts can be  established using a similar method as that used to prove Lemma \ref{lem2}. Let $i\in\mathcal{V}$.

\noindent{F1.} If $s_i(t)\leq \zeta_0 h(0)+ (1-\zeta_0) H(0)$ for some $\zeta_0\in(0,1)$, then $s_i(t+1)\leq \lambda_\ast \zeta_0  h(0)+(1- \lambda_\ast \zeta_0  )H(0) $, where $\lambda_\ast =1-\alpha(n-1)$.

\noindent{F2.} If $s_i(t)\leq \zeta_0 h(0)+ (1-\zeta_0) H(0)$ for some $\zeta_0\in(0,1)$ and $(i,j)\in G_t$, then  $s_j(t+1)\leq  \alpha \zeta_0 h(0)+ (1-\alpha \zeta_0) H(0)$.

By recursively applying F1 and F2, and exploiting the properties of $v_{m_1},\dots,v_{m_{n-1}}$, we obtain
\begin{align}
&\mathds{P}\Big( s_i\big(K_0\big)\leq \frac{\rho_\ast^{K_0}}{2} h(0)+ \big(1-\frac{\rho_\ast^{K_0}}{2}\big)H(0),\ i\in\mathcal{V} \Big) \nonumber\\
&\geq  p_\ast^{n-1} \prod_{t=0}^{K_0-1}\big(b_t(1-d_t)\big). \nonumber
\end{align}
This implies
\begin{align}\label{60}
\mathds{P}\Big( \mathcal{H}\big(K_0\big)\leq  \big(1-\frac{\rho_\ast^{K_0}}{2}\big)\mathcal{H}(0) \Big)\geq  p_\ast^{n-1} \prod_{t=0}^{K_0-1}\big(b_t(1-d_t)\big).
\end{align}
On the other hand, from the definition of the algorithm we know that
\begin{align}\label{61}
\mathds{P}\big( \mathcal{H}\big(t+1\big)\leq  \big(1+2\beta(n-1) \big)\mathcal{H}(0) \big)= 1
\end{align}
and
\begin{align}\label{62}
\mathds{P}\big( \mathcal{H}\big(K_0\big)>\mathcal{H}(0) \big)\leq 1- \prod_{t=0}^{K_0-1}(1-d_t).
\end{align}

Since $\{{B}_t\}_{t\ge 0}$, $\{{D}_t\}_{t\ge 0}$, and $\{G_t\}_{t\ge 0}$ define independent processes, we conclude from (\ref{60}), (\ref{61}), and (\ref{62}) that
\begin{align}
\mathds{E}\big\{\mathcal{H}\big((m+1)K_0\big)\big|\mathcal{H}\big(m K_0\big)  \big\} \leq \big(1- X_m+ Y_m\big)\mathcal{H}\big(m K_0\big). \nonumber
\end{align}
The claim follows directly from Lemma \ref{lem7} and (\ref{61}). \hfill$\blacksquare$

\subsubsection*{Proof of Theorem \ref{thmrsr3}} One can easily see that:
\begin{align}
\mathds{P}\Big(\mathcal{H}(t+1)\geq \big(1-2(n-1)\alpha\big)\mathcal{H}(t)\Big)=1, \nonumber
\end{align}
and
\begin{align}
\mathds{P}\big(\mathcal{H}(t+1)<\mathcal{H}(t)\big)\leq b.\nonumber
\end{align}

Define $L_0:=\inf \{t\in \mathds{Z}: (1+\beta)^t\geq 2(n-1)\}$. Consider time intervals $[mK,(m+1)K-1]$ for $m=0,1,\dots, (n^2-n)(L_0-1)$. Denote $K_{L_0} =K((n^2-n)(L_0-1)+1)$. Under Assumption A8 and based on the fact that there are at most $n(n-1)$ arcs, there are two nodes $i_\ast,j_\ast \in \mathcal{V}$ and $L_0$ instants $0\leq \tau_1<\tau_2<\dots<\tau_{L_0}<K_{L_0}$    such that $(i_\ast,j_
\ast) \in\mathcal{G}^-_{\tau_k}$ and $|s_{i_\ast}(\tau_k)-s_{j_\ast}(\tau_k)|\geq \mathcal{H}(\tau_k)/(n-1)$ for all $\tau_k$.  As a result, we have
\begin{align}
&\mathds{P}\Big(\mathcal{H}(K_{L_0})\geq |s_{i_\ast}(K_{L_0})-s_{j_\ast}(K_{L_0})|\geq \mathcal{H}(0)(1+\beta)^{L_0} \cdot (n-1)^{-1}\Big)\geq \big( d p_\ast(1-p^\ast)^{n-2}\big)^{L_0} (1-b  )^{K_{L_0}}. \nonumber
\end{align}
The previous inequality is obtained by considering the events where $D_{\tau_k}=1$ and $i_\ast= N^-_{j_\ast} (\tau_k)$ for all $\tau_k$ and $B_t=0$ for all $t\in[0,K_{L_0}-1]$.

Then the desired conclusion is obtained by applying the same argument as that used at the end of the proof of Theorem \ref{thmsr3}. This completes the proof. \hfill$\blacksquare$

\subsubsection*{Proof of Proposition \ref{prop2}} The result follows from a similar  sample-path analysis using the second Borel-Cantelli lemma as in the proof of Proposition \ref{prop1}. Hence we omit the details. \hfill$\blacksquare$

\subsubsection*{Proof of Theorem \ref{thmrsr4}} Let us focus on a given positive cluster  $\mathcal{V}^\dag$ of $\mathcal{G}$. Again, we use the following notations
$H^\dag(t)=\max_{i\in \mathcal{V}^\dag} s_i(t)$, $h^\dag(t)=\min_{i\in \mathcal{V}^\dag} s_i(t)$, $\mathcal{H}^\dag(t)= H^\dag(t)-h^\dag(t)$. Applying the analysis of  Lemma \ref{lem10} on $\mathcal{V}^\dag$ we know that $\mathcal{H}^\dag(t)$, $H^\dag(t)$, and $h^\dag(t)$ converge to finite limits almost surely if $\sum_{t\geq 0} d_t<\infty$.

Applying the analysis of Theorem \ref{thmrsr2} on $\mathcal{V}^\dag$, we get
\begin{align}\label{99}
&\mathds{E}\big\{\mathcal{H}^\dag\big((m+1)K_0\big)\big|\mathcal{H}^\dag \big(m(K_0)\big)  \big\} \leq  \big(1-X_m\big)
\mathcal{H}^\dag\big(mK_0\big)+(1+2\beta)\sum_{t=mK_0}^{(m+1)K_0-1} d_t \mathcal{H}(t),
\end{align}
where $X_m$ is defined in Theorem \ref{thmrsr2}. From (\ref{100}) we know that $\mathds{E}(\mathcal{H}(t))\leq  \mathcal{H}_0\prod_{t=0}^\infty\big(1+ 2\beta (n-1)d_t\big)$ for all $t\geq 0$. Taking the expectation on both sides in (\ref{99}), we obtain:
\begin{align}
&\mathds{E}\big\{\mathcal{H}^\dag\big((m+1)K_0\big)\big\} \leq  \big(1-X_m\big)
\mathds{E}\big\{ \mathcal{H}^\dag\big(mK_0\big)\big\}  + \Big[(1+2\beta)\mathcal{H}_0\prod_{t=0}^\infty\big(1+ 2\beta (n-1)d_t\big)\Big]W(m). \nonumber
\end{align}

Moreover, since $\prod_{t=0}^{\infty}(1-d_t)<\infty$, $\sum_{m=0}^\infty J(m)=\infty$ implies $\sum_{m=0}^\infty X_m=\infty$. In view of Lemma \ref{lem7}, we have  $\lim_{m\rightarrow \infty}\mathds{E}\big\{\mathcal{H}^\dag\big((m+1)K_0\big)\big\}=0$ if $\lim_{m \rightarrow \infty} W(m)/J(m)=0$. Invoking Fatou's lemma (e.g., Theorem 1.6.5, \cite{durr}), we further conclude that
$\mathds{E}\big\{\liminf_{m\rightarrow \infty} \mathcal{H}^\dag\big((m+1)K_0\big)\big\}\leq \lim_{t\rightarrow \infty}\mathds{E}\big\{\mathcal{H}^\dag\big((m+1)K_0\big)\big\}=0$. Hence $\mathds{E}\big\{\lim_{m\rightarrow \infty} \mathcal{H}^\dag\big((m+1)K_0\big)\big\}=0$ since $\mathcal{H}^\dag\big((m+1)K_0\big)$ converges almost surely if $\sum_{t\geq 0}d_t<\infty$. Therefore, we have $\mathds{P}\big(\lim_{m\rightarrow \infty} \mathcal{H}^\dag\big((m+1)K_0\big)=0\big)=1$, and $\mathds{P}\big(\lim_{t\rightarrow \infty} \mathcal{H}^\dag\big(t\big)=0\big)=1$. This means that $H^\dag(t)$ and $h^\dag(t)$ converge to the same limit, which must be the limit of the each node state in $\mathcal{V}^\dag$, which completes the proof.
  \hfill$\blacksquare$

\section{Conclusions}
Inspired by various examples from social, biological and engineering networks,  the emerging behaviors of node states  evolving over  signed random networks in a dynamical environment were studied. Each node received  positive and negative  recommendations from its neighbors determined by the sign of the interaction arcs. After receiving recommendations, each node put a deterministic  weight and a random attention on each of the recommendations and then updated  its state.  Various conditions were derived  the almost sure convergence, divergence, and clustering for both state-reversion model and relative-state-reversion model. The results showed explicitly  in general  positive arcs contribute to convergence, negative arcs contribute to divergence, while the structure of the sign patterns contribute to clustering. Some interesting future directions  include the co-evolution of the signs of the interaction links   along with the node states,  as well as the optimal placement of  negative links with the aim of breaking  the effect of positive updates as much as possible.

\end{document}